\documentclass[11pt,twoside]{article}
\usepackage{asp2010}
\resetcounters

\begin{document}

\title{The Gaia-ESO Survey astrophysical calibration}
\author{Elena~Pancino$^1$, on behalf of the Gaia-ESO Survey consortium
\affil{$^1$INAF --- Osservatorio Astronomico di Bologna, Via C. Ranzani 1,
I-40127 Bologna, Italy}}

\begin{abstract}

The Gaia-ESO Survey is a wide field spectroscopic survey recently started with
the FLAMES@VLT in Cerro Paranal, Chile. It will produce radial velocities more
accurate than Gaia's for faint stars (down to V$\simeq$18), and astrophysical
parameters and abundances for approximately 100\,000 stars, belonging to all
Galactic populations. 300 nights were assigned in 5 years (with the last year
subject to approval after a detailed report). In particular, to connect with
other ongoing and planned spectroscopic surveys, a detailed calibration program
--- for the astrophysical parameters derivation --- is planned, including well
known clusters, Gaia benchmark stars, and special equatorial calibration fields
designed for wide field/multifiber spectrographs.

\end{abstract}

\section{The Gaia-ESO Survey}

Gaia-ESO is a public spectroscopic survey, targeting $\geq$10$^5$~stars,
systematically covering all major components of the Milky Way, from halo to star
forming regions, providing the first homogeneous overview of the distributions of
kinematics and elemental abundances. This alone will revolutionise knowledge of
Galactic and stellar evolution: when combined with Gaia astrometry the survey will
quantify the formation history and evolution of young, mature and ancient Galactic
populations. With well-defined samples, it will survey the bulge, thick and thin
discs and halo components, and open star clusters of all ages and masses. The FLAMES
spectra will: quantify individual elemental abundances in each star; yield precise
radial velocities for a 4-D kinematic phase-space; map kinematic gradients and
abundance - phase-space structure throughout the Galaxy; follow the formation,
evolution and dissolution of open clusters as they populate the disc, and provide a
legacy dataset that will add enormous value to the Gaia mission and ongoing ESO
imaging surveys.

\subsection{Scientific aims}

How disc galaxies form and evolve, and how their component stars and stellar
populations form and evolve, are among the most fundamental questions in
contemporary astrophysics \citep{kormendy10,peebles11,komatsu11}. The Gaia-ESO
survey will contribute to those key questions, by revolutionising our knowledge
of the formation and evolution of the Milky Way Galaxy and the stars that
populate it. Because stars form in associations and clusters rather than singly,
understanding star formation in the Milky Way also implies studying cluster
formation.

The key to decoding the history of galaxy evolution involves chemical element
mapping, which quantifies timescales, mixing and accretion length scales, and
star formation histories; spatial distributions, which relate to structures and
gradients; and kinematics, which relates to both the felt but unseen dark
matter, and dynamical histories of clusters and merger events \citep{freeman02}.
With Gaia, and calibrated stellar models, one will also add ages. Manifestly,
very large samples are required to define all these distribution functions and
their spatial and temporal gradients. 

With more than 10$^5$ stars and 100 clusters, each with complete 6D
space mapping when combined with Gaia, and with the addition of astrophysical
parameters (T$_{\rm{eff}}$, log$g$, [M/H]), abundance ratios (iron-peak and
$\alpha$-elements, plus other species for 10\,000 stars), and of ages for
clusters, the Gaia-ESO Survey is the dataset needed to answer those questions.
The expected scientific output is enormous, and a brief summary of the main
survey goals is reported in the following.

\medskip

{\bf Clusters and stellar evolution.} Theories of cluster formation range from
the highly dynamic through to quasi-equilibrium and slow contraction scenarios.
These different routes lead to different initial cluster structures and
kinematics \citep{jackson10}. Whilst hydrodynamic and N-body simulations are
developing, a fundamental requirement is an extensive body of detailed
observations. A complete comparison requires precise position and velocity
phase-space information resolving the internal cluster kinematics
($\leq$0.5~km/s). 

Moreover, each star cluster provides a (near-)coeval snapshot of the stellar
mass function, suitable for testing stellar evolution models from pre-main
sequence phases right through to advanced evolutionary stages. Much of the input
physics in stellar models can be tested by its effects on stellar luminosities,
radii and the lifetimes of different evolutionary phases. Homogeneous
spectroscopy will provide estimates of stellar parameters and reddening for
large samples of stars over a wide range of masses, in clusters with a wide
range of ages and mean chemical compositions. 

\medskip

{\bf The halo and the Bulge.} Recent surveys have revealed that the halos of
both our own and other Local Group galaxies are rich in substructures
\citep{belokurov06}. These not only trace the Galaxy's past, but have enormous
potential as probes of its gravitational field and hence as tracers of the still
very uncertain distribution of dark matter \citep{helmi04}. High precision
radial velocities for many stars at latitudes $|b|>30\deg$ will lead to the
discovery of more substructures. Their abundance patterns will indicate clearly
whether a given structure represents a disrupted object and of which type, or
has formed dynamically by resonant orbit-trapping. The kinematics of streams
will place tight constraints on the distribution of dark matter. 

In simulations of galaxy formation, mergers tend to produce substantial bulges
made of stars that either formed in a disc that was destroyed in a merger, or
formed during a burst of star formation that accompanied the merger
\citep{abadi03}. Such ``classical" bulges are kinematically distinguishable from
``pseudo-bulges" that form when a disc becomes bar unstable, and the bar buckles
into a peanut-shaped bulge \citep{peebles11,kormendy10}. In common with the great
majority of late-type galaxies, the Galaxy's inner bulge appears to be a
pseudo-bulge, but $\rm{\Lambda}$CDM simulations suggest that it should also host
a classical bulge, perhaps that observed at larger radii. By studying the
kinematics and chemistry of K giants at $|b|>5\deg$ we will either confirm the
classical bulge or place limits on it which will pose a challenge to
$\rm{\Lambda}$CDM theory.

\medskip

{\bf The discs.} Thick discs seem common in large spiral galaxies
\citep{gilmore83,yoachim06}. Are they evidence that the last major merger event
occurred very much longer ago than is expected in standard cosmologies? Are they
artifacts of thin disc dynamical evolution? Are they both or neither of these?
How did the metallicity of the ISM evolve at very early times? How does this vary
with Galactocentric distance? Do major infall events occasionally depress the
metallicity of the ISM17? The Gaia-ESO Survey will determine quantitative
kinematics and abundance patterns for large samples of thick disc F ang G stars
over one outer radial and three vertical scale lengths to help elucidate these
key questions in Galaxy formation and evolution. 

The selected sample of $\simeq$5000 F and G stars (see below) within 1~kpc from the
Sun covers both thin and thick discs, and all ages and metallicities. Using field
stars and clusters, where ages are also known, the Gaia-ESO Survey will explore the
region from about 6 to more than 20 kpc in Galactocentric distance, and will trace
chemical evolution as a function of age and Galactocentric radius across a disc
radial scale length. These are key inputs to models for the formation and evolution
of the Galaxy disc. Current estimates suffer from poor statistics, inhomogeneous
abundance determinations and absence of data at key ages and orbits
\citep{nordstrom04}. The Gaia-ESO Survey will also address current disc structure,
that which hosts the star formation. Spiral structure is fundamental to the dynamics
of the disc: it dominates the secular rise in the random velocities of stars, and
may even cause radial migration of stars and gas \citep{antoja10}. Currently, we are
not even clear about the global morphology of our spiral structure, and the
information we have on its dynamics largely relates to gas, not stars. We will
initiate a study of the kinematic distortion in the disc potential due to the
bar/spirals by measuring some 1000s of radial velocities down key arm, inter-arm and
near-bar lines of sight.

\subsection{Survey organization}

\begin{figure}
\plotone{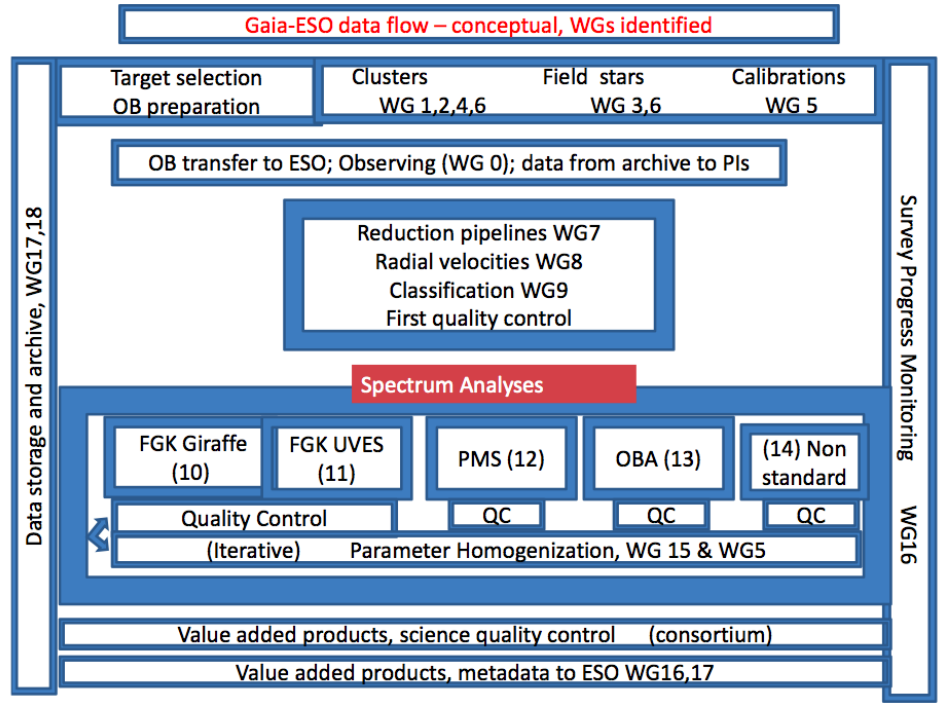}
\caption{An overwiev of the Gaia-ESO Survey data flow system.
\label{pancino1_fig_smp}}
\end{figure}

The survey has approximately 300 co-investigators, and the work is structured in
a series of documents agreed with ESO, principally the Survey Management Plan
and the Survey Implementation Plan. Fig.~\ref{pancino1_fig_smp} shows the work
organization flow, where each WG (Working group) is indicated.

The obtained raw data will become publicly available through the ESO archive as soon
as they are obtained. There will be different advanced data products releases:

\begin{itemize}
\item{semestral data releases: will begin 12 months after observations started
(31 December 2011) and they will refer to all targets that were completed six
months before the release date; they will contain reduced 1D spectra with
variance, radial velocity with uncertainty, basic target information (including
variability);}
\item{annual data releases: they will start 18 months after observations started
and will refer to all targets that were completed six months before the release
data; they will contain astrophysical parameters determination for the single
stars and for the clusters as a whole;}
\item{final data release: containing the full determinable set of astrophysical
parameters for each individual target, and for the open clusters as systems,
with updated and consistent calibration.}
\end{itemize} 

\subsection{Observing strategy}

The Gaia-ESO Survey was awarded 300 observing nights (60 nights per year, with
the last year subject to approval after a progress review) with FLAMES at the
ESO VLT (Very Large Telescope). FLAMES \citep{FLAMES} feeds fibers to two
spectrographs: UVES \citep{UVES}, with a resolution of R$\simeq$47\,000,
receives 8 fibers and GIRAFFE, with a resolution ranging from R$\simeq$15\,000
to 20\,000, receives 132 fibers. Part of the fibers are dedicated to the sky, and
a few special fibers are illuminated by wavelength calibration lamps, allowing
for a radial velocity determination to better than 100~m/s. Observations started
in December 2011.

A selection of the order of 10$^5$ stars belonging to all Galactic components
will be obtained from exisiting photometric surveys such as 2MASS \citep{2MASS},
VISTA \citep{VISTA}, SDSS \citep{SDSS} and from dedicated photometries either
found in the literature \citep[][to name a few]{dias02,kharchenko05} or
specifically derived from public archival data. Observations are restricted to
+10$\deg$$\geq$Dec$\geq$--10$\deg$ whenever possible, to minimize airmass
limits, and to 9$\geq$V$\geq$19~mag (where for V$>$17~mag only radial velocities
will be measured).

The primary targets in the various Galactic components will be:

\begin{itemize}
\item{bulge: $\simeq$10\,000 K giants belonging to the red clump
(I$\simeq$156~mag), for an abundance analysis of iron-peak and $\alpha$-elements
with both UVES and GIRAFFE;} 
\item{halo and thick disk: F and G stars, with 17$\geq$r$\geq$18~mag, for
iron-peak and $\alpha$-elements down to [Fe/H]$\simeq$--1.0~dex; stars belonging
to known streams (e.g., Sgr) will be targeted; the halo targets are expected to
be many thousands, as are the thin+thick disk stars;}
\item{outer thick disk (2--4~kpc fonr the Sun): F and G stars, with 25\% of the
fibers allocated to candidate K giants (r$\leq$18~mag) for studying the warp
and the Monoceros stream;}
\item{thin disk dynamics: six fields at I$\leq$19~mag will target red clump
stars for disk spiral arm/bar dynamics, and only radial velocities will be
obtained;}
\item{Solar neighborhood: UVES parallel observations of approximately 5000 G
stars within 1~kpc from the Sun, for a detailed abundance analysis of all
available elements in the 4800--6800~\AA\  range.}
\item{Open clusters: a total of $\sim$100 clusters of all ages (excluding the
embedded phase) will be observed, choosing high-probability members of all
spectral types --- as appropriated --- from O to K dwarfs and giants, and
including unveiled PMS (Pre-Main Sequence) stars; the faintest targets will
provide accurate radial velocities, the brightest ones a detailed chemical
abundance analysis;}
\item{calibration fields: these are discussed in Sect.~\ref{pancino1_sec_cal};}
\item{archival data: the Gaia-ESO Survey will analyse all ESO archival data
consistent with the observing set-ups and the scientific goals of the survey.}
\end{itemize}

\section{Astrophysical calibration} \label{pancino1_sec_cal}

This conference was focused mainly on the standardization and calibration of
physical quantities, such as the flux or the wavelength, that can be directly
measured, and on the impact of factors that make those measures difficult, such
as the effect of the atmosphere. 

However other astrophysical quantities are derived in a much more indirect way,
by combining direct measurements (for example equivalent width of absorption
lines, or oscillator strengths measured for the corresponding transitions) on
sophisticately treated data with theoretical models (for example stellar
atmospheric models). The resulting astrophysical parameters (T$_{\rm{eff}}$,
log$g$, [Fe/H], [$\alpha$/Fe], and other abundance ratios) --- that will be
derived in the Gaia-ESO Survey --- also need their own calibration. However, an
astrophysical calibration is based on the comparison of measurements that are
often of comparable quality to each other, and ultimately can be described as
the effort of estimating the systematic uncertainties underlying a set of
indirect measurements. 

\begin{figure}
\centering
\includegraphics[width=9cm]{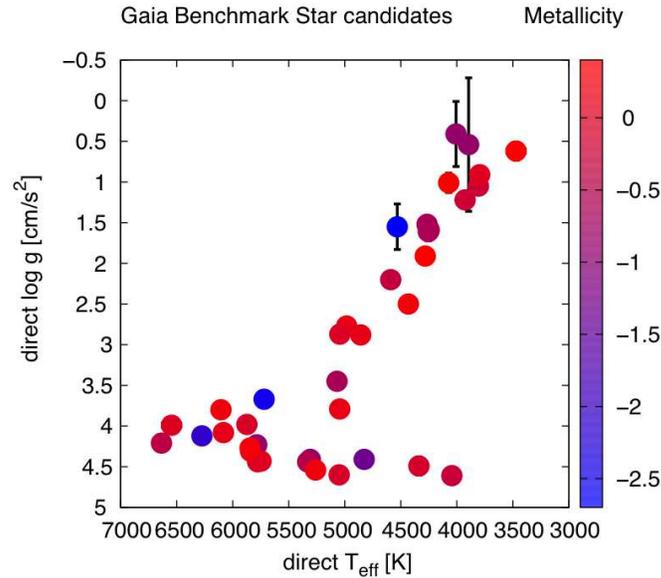}
\caption{The Gaia benchmark stars, with their direct measurements of
T$_{\rm{eff}}$ and log$g$ (courtesy of U.~Heiter).
\label{pancino1_fig_gbs}}
\end{figure}

The basic example of astrophysical calibrator in the case of high-resolution
abundance analysis of stellar atmospheres, is the Sun. It is studied with a much
higher precision, with much better data (because it is extremely bright) and by
many different groups with different technicques. Thus, all astronomers deriving
an abundance analysis of solar type stars, analyze the Sun as well with the same
method, and compare their results with the consensus solar abundance set.
However, the Sun is a good calibrator only for solar metallicity dwarfs, and
there is of course no guarantee that it will be as good for metal-poor giants,
for example.  

Thus, the Gaia-ESO survey dedicates a fraction of the time (approximately 100~h)
to observations of calibrators with various purporses. These are:

\begin{itemize}
\item{a selection of stars for radial velocity calibration mainly from
\citet{crifo10}, a large catalogue of stars which are stable to 300~m/s and that
will be used for the radial velocity calibration of Gaia as well;}
\item{a selection of stars covering the parameters space of the Milky Way field
pointings (Fig.~\ref{pancino1_fig_gbs}); these have their parameters determined
independently and as directly as possible (i.e., using parallaxes, diameter
measurements and so on) and are also used to calibrate the parametrizer
algorithms of the Gaia pipelines;}
\item{a selection of templates peculiar stars of types which may fall into our
selection windows, including barium and carbon stars, r- and s-process enhanced
stars \citep[e.g., from][]{alksnis01};}
\item{a set of approximately 30--50 globular and open clusters, which will cover
the entire metallicity scale covered by the Gaia-ESO Survey; these are chosen as
external calibrators among the best studied clusters in the literature, and
contain objects in common with other ongoing or planned spectroscopic surveys
\citep[e.g., RAVE, HERMES, APOGEE, see also][]{lane11,frinchaboy10};}
\item{a set of internal calibrators, typically relatively young open clusters,
containing stars of different spectral types, to link different abundance
analysis methods, such as those employed for hot stars (O, B, and A), cool stars
(F, G, K, and even M stars), or PMS stars of all spectral types; nearby open and
globular clusters will ensure the link between dwarfs and giants;}
\item{a set of pre-defined fields around the celestial equators and possibly at
the Ecliptic Pole, containing a mix of objects with different characteristics, to
be used as calibrating fields for stellar spectroscopic surveys carried out with
wide field multi-object spectrographs; two of these fields are in the areas
surveyed by the Corot mission, in the direction of the Galactic center and
anti-center \citep{gazzano10}.}
\end{itemize}

Thus, the Gaia-ESO Survey will maximise its legacy value by providing all the
tools to link its measurements with past and future studies, ultimately with the
goal of combining all the large stellar spectroscopic sturveys together.


\bibliography{Pancino_E_1}

\end{document}